# Current Renormalisation Constants with an $O(a)$-improved Fermion Action

UKQCD Collaboration

D.S. Henty, R.D.Kenway, B.J. Pendleton, J.I. Skullerud

*Department of Physics and Astronomy, The University of Edinburgh, Edinburgh EH9 3JZ, Scotland*

## Abstract

Using chiral Ward identities, we determine the renormalisation constants of bilinear quark operators for the Sheikholeslami-Wohlert action lattice at $\beta = 6.2$. The results are obtained with a high degree of accuracy. For the vector current renormalisation constant we obtain $Z_V = 0.817 \pm 2 \pm 8$, where the first error is statistical and the second is due to mass dependence of $Z_V$. This is close to the perturbative value of 0.83. For the axial current renormalisation constant we obtain $Z_A = 1.045 ^{+10}_{-14}$, significantly higher than the value obtained in perturbation theory. This is shown to reduce the difference between lattice estimates and the experimental values for the pseudoscalar meson decay constants, but a significant discrepancy remains. The ratio of pseudoscalar to scalar renormalisation constants, $Z_P/Z_S$, is less well determined, but seems to be slightly lower than the perturbative value.

hep-lat/9412088 19 Dec 94

Typeset using REVTEX

# I. INTRODUCTION

In a recent paper [1], we reported values for the decay constants $f_\pi$ and $f_K$ from lattice calculations that were considerably lower than the experimental values. This has been a persistent feature of lattice calculations of pseudoscalar decay constants [2,3]. However, it has not been possible to attribute this discrepancy to any particular systematic error in the calculations (eg quenching), because of uncertainties in the renormalisation of the current operators involved.

A fully non-perturbative determination of renormalisation constants for finite operators can be achieved using chiral Ward identities [4,5], thereby bypassing the need for perturbative calculation of these quantities. A preliminary calculation at $\beta = 6.0$, presented in [5], indicated that the axial vector current renormalisation, in particular, differs significantly from its perturbative value. In this paper, we present results for the vector, axial-vector, pseudoscalar and scalar renormalisation constants at $\beta = 6.2$, and update the values for decay constants given in [1].

## II. LATTICE WARD IDENTITIES AND RENORMALISATION CONSTANTS

We define the (non-conserved) lattice vector current, the axial current and the pseudoscalar and scalar densities as follows:

$$V_\mu^{La}(x) = \overline{\psi}(x)\gamma_\mu \frac{1}{2}\lambda^a \psi(x) \tag{1}$$

$$A_\mu^{La}(x) = \overline{\psi}(x)\gamma_\mu\gamma_5 \frac{1}{2}\lambda^a \psi(x) \tag{2}$$

$$P^a(x) = \overline{\psi}(x)\gamma_5 \frac{1}{2}\lambda^a \psi(x) \tag{3}$$

$$S^a(x) = \overline{\psi}(x)\frac{1}{2}\lambda^a \psi(x) \tag{4}$$

In the continuum limit, these operators are related to operators obeying the correct current algebra by multiplicative renormalisation constants $Z_V^L$, $Z_A^L$ etc, so that $V_\mu = Z_V^L V_\mu^L$ etc. [5].

The renormalisation constants for $V_\mu^L$ can easily be determined by evaluating

$$Z_V^L = \frac{\sum_{\vec{x}}\langle P^\dagger(\vec{0},0)P(\vec{x},T)\rangle}{\sum_{\vec{x},\vec{y}}\langle P^\dagger(\vec{0},0)V_4^L(\vec{y},t)P(\vec{x},T)\rangle} \tag{5}$$

Inserting a complete set of states and noting that the matrix element of $V_4^L(0)$ between a degenerate pseudoscalar meson state $P_n$ is $\langle P_n \,|\, V_4^L \,|\, P_n \rangle = 2E_n/Z_V^L$ , we see that this should give a precise estimate provided the effect of the off-diagonal matrix elements $\langle P_m \,|\, V_4^L \,|\, P_n \rangle$ can be neglected.

For the axial case, there is no conserved current, or any other "easy" way of determining the renormalisation constants, but they can be obtained using chiral Ward Identities. Using the arguments of [4,5], we obtain the following identities for the Sheikholeslami–Wohlert (SW) [6] action:

$$2\rho \sum_{x,\vec{y}} \langle P^a(x) A_\nu^{Lb}(y) V_\rho^{Lc}(0) \rangle$$

$$= -i(\frac{Z_V^L}{Z_A^{L\,2}} - \rho r a) f^{abd} \sum_{\vec{y}} \langle V_\nu^{Ld}(y) V_\rho^{Lc}(0) \rangle$$

$$-i(\frac{1}{Z_V^L} - \rho r a) f^{acd} \sum_{\vec{y}} \langle A_\nu^{Lb}(y) A_\rho^{Ld}(0) \rangle \qquad (6)$$

and

$$2\rho \sum_{x,\vec{y}} \langle P^a(x) S^b(y) P^c(0) \rangle$$

$$= (\frac{Z_P}{Z_A^L Z_S} - \rho r a) d^{abd} \sum_{\vec{y}} \langle P^d(y) P^c(0) \rangle$$

$$(\frac{Z_S}{Z_A^L Z_P} - \rho r a) d^{acd} \sum_{\vec{y}} \langle S^b(y) S^d(0) \rangle \qquad (7)$$

where

$$\rho = \frac{\langle 0 \mid \partial_4 A_4^{La} \mid P^a \rangle}{2 \langle 0 \mid P^a \mid P_a \rangle} \qquad (8)$$

and $d^{abc}$ is defined by

$$\{\lambda^a, \lambda^b\} = 2d^{abc}\lambda^c + \frac{4}{N_f}\delta^{ab} \qquad (9)$$

## III. COMPUTATION AND RESULTS

We have performed simulations at $\beta = 6.2$ on a $24^3 \times 48$ lattice. We have generated propagators using the SW action, with two quark masses, corresponding to $\kappa = 0.14144$ and $0.14262$, where $\kappa = 1/2(m_0 + 4r)$ and $r = 1$. 60 configurations have been analysed at $\kappa = 0.14144$, and 26 configurations at $\kappa = 0.14262$ (for details, see [7]). The statistical errors are calculated with a bootstrap procedure, using 100 bootstrap samples.

$Z_V$ was determined from eq. (5), using 10 configurations, at three values for the quark mass (corresponding to $\kappa = 0.14144$, $\kappa = 0.14226$ and $\kappa = 0.14262$). The results are presented as a function of $t$ in fig.1. We see that the values for $Z_V$ are roughly independent of $t$. Our best values, obtained by fitting to timeslices 5–19, are given in table I. The errors from the variation between the timeslices are obtained from fits to 100 bootstrap samples of timeslices within the fit range.

The results are plotted as a function of the square of the mass of the pseudoscalar meson (proportional to the quark mass) in fig.2. We see that the results show a clear (linear) dependence on the quark mass, consistent with the expectation that the leading corrections to our calculations should be of $O(\alpha_s m_0 a)$. Perturbation theory at one-loop level [8,9] with a "boosted" coupling constant [10] gives $Z_V^L \approx 0.83$, which is quite close to our non-perturbative values.

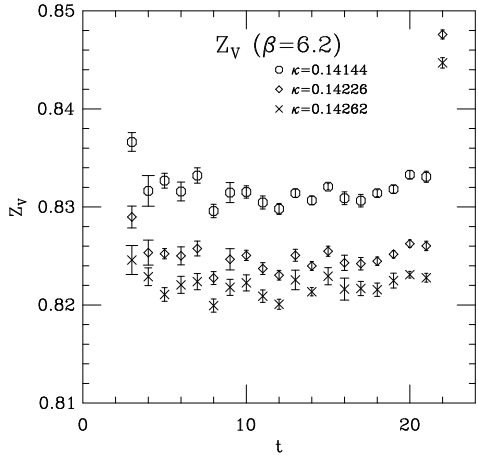

FIG. 1. $Z_V^L$ as a function of $t$.

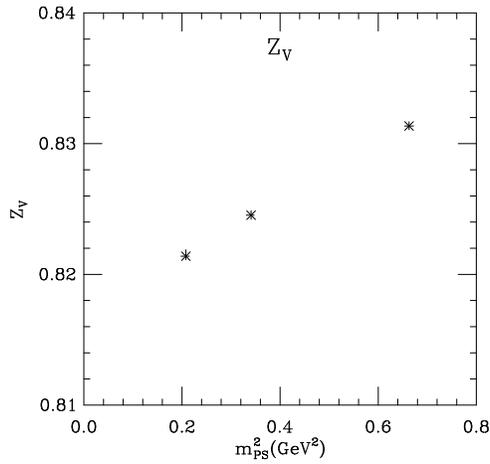

FIG. 2. $Z_V^L$ as a function of the mass of the pseudoscalar meson. The scale is taken from the string tension [7].

| $\kappa$ | $m_{PS}^2$ | $Z_V^L$ |
|---|---|---|
| 0.14262 | 0.208 | $0.82139^{+41+25}_{-12-25}$ |
| 0.14226 | 0.341 | $0.82453^{+24+24}_{-22-23}$ |
| 0.14144 | 0.663 | $0.83136^{+23+22}_{-16-23}$ |

TABLE I. Values of the renormalisation constant $Z_V^L$ as a function of the quark mass. The first set of errors are the statistical errors, while the second set are the errors due to the variation between the timeslices.

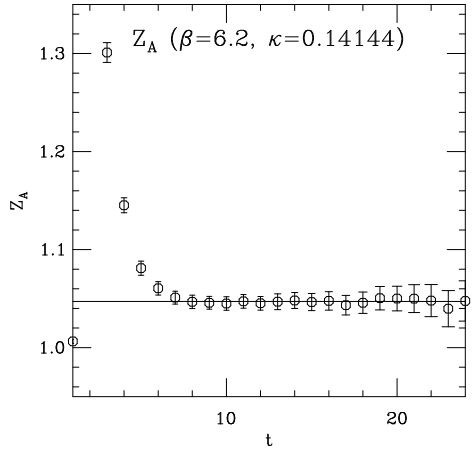

FIG. 3. $Z_A^L$ as a function of $t$ for $\kappa = 0.14144$

| $\kappa$ | $m_{PS}^2$ | $Z_A^L$ | $Z_P/Z_S$ |
|---|---|---|---|
| 0.14262 | 0.208 | $1.040^{+10}_{-9}$ | $0.693^{+28}_{-40}$ |
| 0.14144 | 0.663 | $1.047^{+8}_{-8}$ | $0.649^{+9}_{-8}$ |

TABLE II. Values of the renormalisation constants $Z_A^L$ and $Z_P/Z_S$ as functions of the quark mass.

We have also used eq. (6), with $\nu = \rho = 0$, to check the consistency of our results at $\kappa = 0.14144$, using the value for $Z_V^L$ quoted in table I and the value for $Z_A^L$ given in table II as input. This gives $Z_V^L = 0.817^{+8}_{-10}$, which is within $2\sigma$ of the result obtained from eq. (5).

The axial vector renormalisation constant $Z_A^L$ is determined using eq. (6), with $\nu = \rho = i$ and summing over $i = 1, 2, 3$, using the values for $Z_V^L$ quoted in table I as input. The results for $\kappa = 0.14144$ are plotted against $t$ in fig.3. We see that, apart from the effect of the contact terms on the first few timeslices, they show virtually no dependence on $t$. Our best estimates are given in table II.

Within the statistical errors, these results show no dependence of $Z_A$ on the quark mass. The comparison with results from perturbation theory is more interesting: one-loop calculations with a "boosted" coupling constant give $Z_A \approx 0.97$, which is considerably lower than our non-perturbative results. The discrepancy is higher at lower $\beta$, as expected; in [5] $Z_A$ at $\beta = 6.0$ was found to be 1.09.

In table III we show how the values for the decay constants reported in [1] change when we use the results given above for the renormalisation constants. For $Z_V$, we have extrapolated the values in table I to the limit of zero quark mass, giving $Z_V = 0.817 \pm 2$, with an additional uncertainty due to the quark mass dependence of $Z_V$ of $\pm 0.008$, which corresponds the difference between the value at our largest quark mass and the value for zero quark mass. For $Z_A$, we have taken a best estimate, combining our results at the two $\kappa$-values, of $Z_A = 1.045^{+10}_{-14}$, with the errors corresponding to the spread between the highest and lowest estimate. We see that all the decay constants move closer to the experimental values, but that a significant discrepancy still remains, especially for $f_\phi$ and $f_K$. $f_\pi$ turns out to be about $3\sigma$ away from its experimental value. The APE collaboration has found $f_\pi/(m_\rho Z_A) = 0.186(20)$ at $\beta = 6.2$ [11], which gives a value for $f_\pi/m_\rho$ compatible with experiment.

| | old estimates [1] | updated estimates | experiment |
|---|---|---|---|
| $f_\pi$ | $102^{+6}_{-7}$ MeV | $110^{+7}_{-8}$ MeV | 132 MeV |
| $f_K$ | $123^{+5}_{-6}$ MeV | $133^{+7}_{-7}$ MeV | 160 MeV |
| $1/f_\rho$ | $0.316^{+7}_{-13}$ | $0.311^{+7+1}_{-13-1}$ | 0.28 |
| $1/f_{K^*}$ | $0.298^{+5}_{-9}$ | $0.293^{+5+1}_{-9-1}$ | |
| $1/f_\phi$ | $0.280^{+3}_{-6}$ | $0.276^{+3+1}_{-6-1}$ | 0.23 |
| $f_\pi/m_\rho$ | $0.138^{+6}_{-9}$ | $0.149^{+6}_{-10}$ | 0.172 |
| $f_K/m_\rho$ | $0.160^{+7}_{-8}$ | $0.172^{+8}_{-9}$ | 0.208 |
| $f_K/m_{K^*}$ | $0.144^{+4}_{-6}$ | $0.155^{+5}_{-7}$ | 0.179 |

TABLE III. Values of decay constants in physical units, using perturbative and non-perturbative values for the renormalisation constants. The second set of errors in the vector meson decay constants are systematic uncertainties due to the quark mass dependence of $Z_V$.

The ratio of pseudoscalar to scalar renormalisation constant is determined using (7). The results are given in table II. The uncertainty in these results is too large to determine whether there is any dependence on the quark mass here. Perturbative calculations with a "boosted" coupling constant give $Z_P/Z_S = 0.68$. As can be seen, our result for the heavier quark mass (which is the more accurate) is slightly lower than this, while the lighter quark mass gives a value compatible with perturbative results (although the errors here are still quite large). Comparison with the result reported in [5] at $\beta = 6.0$ shows that, as in the case with $Z_A$, the discrepancy decreases with increasing $\beta$.

## IV. CONCLUSIONS

In this paper we have reported the determination of lattice renormalisation constants using chiral Ward identities with the Sheikholeslami-Wohlert action. Our results are obtained with good accuracy, yielding values for $Z_V$ close to the values from perturbation theory (but increasing with increasing quark mass), while the value for $Z_A$ is considerably higher than perturbative results. For $Z_P/Z_S$ the uncertainties are larger, but the results we can have confidence in lie slightly lower than perturbative values. The result for $Z_A$ brings our estimates for $f_\pi$ and $f_K$ considerably closer to experimental values — within $3\sigma$ for $f_\pi$.

## ACKNOWLEDGMENTS

This work was supported by SERC grants GR/J21347 and GR/J98202 and carried out on a Meiko i860 Computing Surface supported by SERC grant GR/G32779, Meiko Limited, and the University of Edinburgh. JIS acknowledges the support of the Norwegian Research Council. DSH akcnowledges the support of PPARC through a Personal Fellowship. We are grateful to David Richards for help with the code, and to Chris Sachrajda for valuable discussions.